\documentstyle[preprint,prl,aps]{revtex}

\def\beq{\begin{equation}}
\def\eeq{\end{equation}}
\def\beqa{\begin{eqnarray}}
\def\eeqa{\end{eqnarray}}

\def\rra{\right\rangle}

\def\lsim{\mathrel{\raise.3ex\hbox{$<$\kern-.75em\lower1ex\hbox{$\sim$}}} }
\def\gsim{\mathrel{\raise.3ex\hbox{$>$\kern-.75em\lower1ex\hbox{$\sim$}}} }

\begin{document}

\preprint{{\vbox{\hbox{UR-1576} \hbox {Jul 1999}
\hbox {rev. Oct 1999} }}}
\draft


\title{On Neutrino Masses and Mixings from Extra Dimensions}
\author{\bf Ashok Das and Otto C.~W.~Kong
\footnote{Present address: Insitute of Physics, Academia
Sinica, Taipei, Taiwan 11529. 
E-mail: kongcw@phys.sinica.edu.tw
.}
}
\address{Department of Physics and Astronomy,\\
University of Rochester, Rochester NY 14627-0171}
\maketitle

\begin{abstract}
In the framework of a Kaluza-Klein type theory with  the Standard
Model fields localized on a 4-dimensional section while gravity propagates
in a full $4+\delta$-dimensional space-time, we examine a mechanism of
naturally small neutrino mass generation through couplings of the
Standard Model 
singlet fermion(s) living also in the full space-time. A numerical study is 
carried out on the charged current universality constraint from the ratio of 
pion decay partial widths. The bounds obtained on the fundamental mass scale could be stringent. 
\end{abstract}



\section{introduction}
Interests in Kaluza-Klein (KK) type theories with extra space-time 
dimensions has recently been revived with an odd twist --- the usual Standard
Model (SM) fields with its chiral fermionic content are assumed to be
localized on a 4-dimensional section. With only gravity propagating
in the full $4+\delta$-dimensional space-time, the extra dimensions could
be ``large",  with a fundamental mass scale $M_*$ as low as a 
TeV\cite{ADD,ADD2}. The scenario seems to promise
rich phenomenological features, which has been the
subject of many recent studies\cite{pheno}. However, present collider
bounds on $M_*$ are typically in the TeV range, while it has been
shown that astrophysical and cosmological processes provide much
stronger bounds\cite{ADD2,ascos}, rendering collider phenomenology
uninteresting. For $\delta=2$, the best bound is about 
$150\;\mbox{TeV}$; while it is about $30\;\mbox{TeV}$ for $\delta=3$.
As collider experiments are unlikely  to yield information on the
feasibility of the specific KK scenario, it would be interesting to have
other probes into the extra dimensions. Here in this letter, we illustrate
how neutrino physics may just provide us with that.

The problem of neutrino mass generation under such a scenario was 
discussed in Refs.\cite{ADDM,DDG}. The relatively small $M_*$ value
invalidates the popular seesaw mechanism and its various variations.
However, as the so-called right-handed neutrino is a SM singlet fermion,
it may very well live in the full $4+\delta$ dimensions along with gravity.
Naturally small Yukawa couplings to the SM neutrinos can then 
result from a volume factor. The suppression, given by $M_*/M_{Pl}$
is roughly in the right range to account for neutrino oscillations\cite{osc}.
In this letter, we present a careful study of the neutrino mass generation
as well as stringent bounds on $M_*$ obtained from charged current
universality.

\section{A Singlet fermion in 5 Dimensions}
We consider here, for illustrative purposes only, the simple case of one SM
singlet fermion $\Psi$ in a 5-dimensional theory, the latter with co-ordinates
$(x^{\mu},y)$, with $\mu= 0, \cdots, 3$ and the $y$ direction
compactified on a circle of circumference $2 \pi R$ by making the
periodic identification $y \sim y + 2 \pi R$. Conventional SM fields
are restricted  to 
live on the 4-dimensional section at $y=0$. We adopt the
representation where the 5-D gamma matrices are given by
\begin{equation}
\Gamma^{\mu} = \left( \begin{array}{cc} 0 & \sigma^{\mu} \\
\bar{\sigma}^{\mu}
& 0 \end{array} \right) ,\quad  
\Gamma^4 =  \left( \begin{array}{cc} i & 0\\
0 & -i \end{array} \right)\; .
\end{equation}
This matches with the 4-D chiral representation in Ref.\cite{PS} for
$\Gamma^{\mu} = \gamma^{\mu}$ and $\Gamma^4=i\gamma^5$.
The fermion $\Psi$ is a 4-spinor which decomposes as
\begin{equation}
\Psi = \left(\begin{array}{c} i \psi_{\scriptscriptstyle\! L} \\ 
\psi_{\scriptscriptstyle\! R}  \end{array} \right)
\end{equation}
where each component has a 4-D chirality marked by its eigenvalue
under $\gamma^5$; and a phase $i$ is introduced for later convenience. 
When compactifying the fifth dimension, the  Fourier expansion
\begin{equation} \label{fe}
\Psi (x,y) = \sum_{n=-\infty}^{\infty} \frac{1}{\sqrt{2 \pi R}} 
\Psi_n (x) e^{iny/R} \; ,
\end{equation}
applies componentwise, with the resultant 2-spinors 
$\psi_{\scriptscriptstyle\!L\!n}$ and $\psi_{\scriptscriptstyle\!R\!n}$ 
liable to be interpreted as KK towers of independent 4-D Weyl spinors.
From the $y$-component of the 5-D  free action 
$\bar{\Psi}\,i\Gamma^{\alpha}\partial_{\alpha} \Psi$, we have 
\begin{eqnarray}
S_{\Psi} &=& - \int d^4 x dy \bar{\Psi} \gamma^5\partial_y \Psi
\nonumber \\ &=& 
\int d^4 x \frac{n}{R}\left( \psi_{\scriptscriptstyle\!L\!n}^{\dag} 
\psi_{\scriptscriptstyle\!R\!n}^{~} +  \psi_{\scriptscriptstyle\!R\!n}^{\dag} 
\psi_{\scriptscriptstyle\!L\!n}^{~} \right) \; . \label{mnr}
\end{eqnarray}

Next, we consider admissible interactions of the components of $\Psi$
with the SM fields living at $y=0$.
Recall that the SM  leptonic doublet $L$ is a Weyl (2-)spinor, 
conventionally taken to be left-handed. Here we keep careful track
of the spinor structures by sticking to 2-spinor notations and paying
attention to the difference between  $\psi$ and  $\psi^{\dag}$. 
We have
\begin{equation} \label{Sint}
S^{\mbox{int}} = \int d^4 x  \left(\frac{ \lambda_\nu }{\sqrt{M_*}}
 \psi_{\scriptscriptstyle\! R}^{\dag}(x,y\!=\!0) 
\phi^*(x) L(x)  + \frac{ \lambda_\nu^{'} }{\sqrt{M_*}} 
\psi_{\scriptscriptstyle\! L}^{{\scriptscriptstyle c}\dag}(x,y\!=\!0) 
\phi^*(x) L(x)  + \mbox{h.c.} \right) \; ,
\end{equation}
where the factor $\sqrt{M_*}$ is introduced to make the Yukawa
couplings $\lambda$ and $\lambda'$ pure numbers; $\phi$ denotes the
Higgs scalar doublet; and
the superscript $c$ indicates a 2-spinor obtained from the charge 
conjugation of the original 2-spinor. In particular,
$\psi_{\scriptscriptstyle\! L}^{\scriptscriptstyle c}=i\sigma^2
\psi_{\scriptscriptstyle\! L}^*$, and transforms as a right-handed 2-spinor.
Note that we violate the conventional notation here\cite{R}. 

With electroweak symmetry breaking, the above interaction generates
4-D quadratic fermionic terms of the same form
\begin{equation} \label{mm'}
m \psi_{\scriptscriptstyle\!R\!n}^{\dag}  \nu_{\scriptscriptstyle\! L}
+ m^{'} \psi_{\scriptscriptstyle\!L\!n}^{{\scriptscriptstyle c}\dag}  
\nu_{\scriptscriptstyle\! L}
\end{equation}
for each $n$. From Eqs.(\ref{mnr}) and (\ref{mm'}), we obtain a
Majorana mass matrix for the effective 4-D neutral fermions (neutrinos)
of the form 
\beq
{\cal L}_{\rm mass} = \frac{1}{2} N^c {\cal M} N^T 
\eeq
 given by
\small \beq \label{Mfull}
\cal{M} =  \left(\begin{array}{cccccccc} 
0 & m & m' & m & m' & m & m' & \cdots \\ 
m & 0 & 0 & 0 & 0 & 0 & 0 & \cdots \\ 
m' & 0 & 0 & 0 & 0 & 0 & 0 & \cdots \\ 
m & 0 & 0 & 0 & 1/R & 0 & 0 & \cdots \\ 
m' & 0 & 0 & 1/R & 0 & 0 & 0 & \cdots \\ 
m & 0 & 0 & 0 & 0 & 0 & 2/R & \cdots \\ 
m' & 0 & 0 & 0 & 0 & 2/R & 0 & \cdots \\ 
 \vdots & \vdots & \vdots & \vdots & \vdots & \vdots & \vdots &
\ddots \end{array} \right)
\eeq \normalsize
with 
\begin{eqnarray*}
N&=&(\nu_{\scriptscriptstyle\! L}, 
\psi_{\scriptscriptstyle\! R\!0}^{\scriptscriptstyle c},
\psi_{\scriptscriptstyle\! L\!0}^{~},
\psi_{\scriptscriptstyle\! R\!1}^{\scriptscriptstyle c},
\psi_{\scriptscriptstyle\! L\!1}^{~},
\psi_{\scriptscriptstyle\! R\!2}^{\scriptscriptstyle c},
\psi_{\scriptscriptstyle\! L\!2}^{~}, \cdots) \\
N^c&=&(\nu_{\scriptscriptstyle\! L}^{{\scriptscriptstyle c}\dag}, 
\psi_{\scriptscriptstyle\! R\!0}^{\dag},
\psi_{\scriptscriptstyle\! L\!0}^{{\scriptscriptstyle c}\dag},
\psi_{\scriptscriptstyle\! R\!1}^{\dag},
\psi_{\scriptscriptstyle\! L\!1}^{{\scriptscriptstyle c}\dag},
\psi_{\scriptscriptstyle\! R\!2}^{\dag},
\psi_{\scriptscriptstyle\! L\!2}^{{\scriptscriptstyle c}\dag},
 \cdots) \; . \end{eqnarray*}
Notice that we have used general relations of the form
\beq
\psi_{\scriptscriptstyle \!1}^{{\scriptscriptstyle c}\dag}
\psi_{\scriptscriptstyle \!2}^{\scriptscriptstyle c} =
\psi_{\scriptscriptstyle \!2}^{\dag}\psi_{\scriptscriptstyle \!1}^{~}\; .
\eeq
The part of the above mass matrix involving the $n$-th KK states has a
universal structure which allows us to introduce the following
compressed  notation,
\small \beq \label{Mcom}
\cal{M} =  \left(\begin{array}{ccccc} 
0 & m & m' & m & m'  \\ 
m & 0 & 0 & 0 & 0  \\ 
m' & 0 & 0 & 0 & 0 \\ 
m & 0 & 0 & 0 & n/R  \\ 
m' & 0 & 0 & n/R & 0 
\end{array} \right) \; .
\eeq \normalsize
Here, it is to be understood that the columns and rows with the $n/R$ represents 
a repeated structure for each $n$; the latter is always restricted to be non-zero, but
have both signs. The notation can be used for the more general cases with more
dimensions, to be discussed below.

Assuming a hierarchy $m,m^{'}\ll 1/R$,
it is straight forward to work out through a block perturbative analysis
the $``$seesaw" contributions of the heavy $n(\ne 0)$ states to the first 
three (light) states of $\cal{M}$. We have, from the compressed
notation of $\cal{M}$ above, the simple result
\beq 
- \left(\begin{array}{ccc} 
\frac{2mm^{'}}{n/R} & 0 & 0  \\ 
 0 & 0 & 0 \\ 
 0 & 0 & 0  
\end{array} \right) \; .
\eeq
It is easy to see that for the full $\cal{M}$ matrix, as explicitly given by
Eq.(\ref{Mfull}), contributions from each $n$ simply adds up to give the only non-zero 
entry as $\sum_n\frac{2mm^{'}}{n/R}$.  It indicates that only the
(Majorana) mass of 
$ \nu_{\scriptscriptstyle\! L}$, the SM neutrino, receives
perturbative contributions,
which {\it is} what one should expect from the mixing terms. The
perturbative effect from each $n$ is small and the perturbative
assumption seems not to be upset, yet the total sum goes as $\sum_n 1/n$
which diverges naively as $ln\; n$, suggesting that the neutrino mass becomes
infinite!  Actually, however, as the Fourier expansion in
Eq.(\ref{fe}) contains both positive as well as {\it negative} $n$ modes, the
$\sum_n\frac{2mm^{'}}{n/R}$ is always zero (There is no $n=0$ term in
the sum). As we will show explicitly below,
in a somewhat more complicated scenario, a simple perturbational analysis gives
a first order correction to the light mass value $m$ of order 
$ \sum_n \frac{m^2 R^2}{n^2}$, the same form as the case with a
pure Dirac mass matrix discussed in Ref.\cite{ADDM}.  The 
latter corresponds to setting  $m^{'}=0$. 
It is important to note here that with both 
$m$ and $m'$ non-zero, the neutrino mass matrix cannot be consistently written in  
the Dirac form.  The sum to infinity of $\frac{1}{n^2}$
is still divergent for $\delta\leq 2$, but will be well-behaved when
properly truncated.

Note that when the compactification is performed on a $Z_2$ orbifold, about half of the
KK modes are projected out. The surviving $\psi_{\scriptscriptstyle\!R\!n}$ modes are to be 
taken as the linear combinations of the positive and negative $n$ modes for each $n$
which are even under a $y$-reflection; while the surviving $\psi_{\scriptscriptstyle\!L\!n}$ 
modes are the corresponding linear combinations which are odd under the  $y$-reflection.
In particular, the $\psi_{\scriptscriptstyle\! L\!0}$ would be projected out and hence 
decoupled in $\cal{M}$. The surviving $\psi_{\scriptscriptstyle\!L\!n}$ modes 
actually have zero amplitude at the SM world ($y=0$) and $m'$ has to be set to zero.

The admission of both $m$ and $m'$ [both terms in Eq.(\ref{Sint})] is interesting
when more than one family of SM neutrinos are considered. The two 2-spinors in
$\Psi$ can then give mass to two neutrinos with mixing. Let us now consider such
masses and mixings for $\nu_{\!e}$ and $\nu_{\!\mu}$.
Eq.(\ref{Mcom}) is modified as $\cal{M}$ is extended to include a second 
$\nu_{\scriptscriptstyle\! L}$. It then takes the block form
\beq \label{Mbk}
\cal{M} =  \left(\begin{array}{ccc} 
0 & \Omega_o^{T} & \Omega_o^{T}   \\ 
\Omega_o & 0 & 0   \\ 
\Omega_o & 0 & M_n   
\end{array} \right) \; ,
\eeq 
where $\Omega_o$ is a general $2\times 2$ matrix representing mass terms of the
form given in Eq.(\ref{mm'}), but now with  two $\nu_{\scriptscriptstyle\! L}$'s, and 
$M_n =  \left(\begin{array}{cc}    
 0 & n/R  \\ n/R  & 0
\end{array} \right)$. 
Skipping the algebraic details, we simply note that a unitary
transformation can bring $\cal{M}$ to the form
\beq \label{Md}
\cal{M} =  \left(\begin{array}{ccc} 
 \left(\begin{array}{cc} m_1 & 0   \\ 0 & m_2 \end{array} \right) & 0 & \Omega^T \\
 0 & \left(\begin{array}{cc} -m_1 & 0   \\ 0 & -m_2 \end{array} \right)  & -  \Omega^T \\
 \Omega & -\Omega & \left(\begin{array}{cc} n/R & 0   \\ 0 & -n/R \end{array} \right) 
\end{array} \right)\; ,
\eeq
where $\Omega$ may be taken as a general $2\times 2$ matrix, and $m_1$ and $m_2$
are the eigenvalues of $\Omega_o$ taken as a Dirac mass matrix. Our compressed notation
involving the $n \ne 0$ KK modes is still effective. Note that each pair of the light 
states having equal and opposite mass eigenvalues can be considered as a Dirac state
consisting  of a $\nu_{\scriptscriptstyle\! L}$ state and a linear combination of
the zero modes from $\Psi$; each of the pair of (Majorana) states is
then  an  equal admixture
of the latter.

Entries in $\Omega$ in the above form of $\cal{M}$ are of the same order as $m_1$ and
$m_2$ or that of the entries in $\Omega_o$, hence $\ll 1/R$. The diagonal entries of
$\cal{M}$ are, therefore, the approximate mass eigenvalues. We can perform a direct 
perturbative analysis to find out the first order corrections to the light masses,
as well as the mixings between the $n \ne 0$ KK modes and the light states, hereafter
denoted by $\left|\nu_i^\pm \rra$ 
(for diagonal mass entry $\pm m_i \;,\; i=1\;\mbox{or}\;2$).
Let $U$ denote the unitary transformation that diagonalizes $\cal{M}$ in Eq.(\ref{Md}).
Perturbation analysis gives, in our compressed notation, 
\beqa \label{U}
U_{\scriptscriptstyle \!\!+\!i,\pm\!n} &=& 
 \frac{- \omega_{\scriptscriptstyle \!\pm\! i}}{\pm n/R - m_i} \nonumber \\
U_{\scriptscriptstyle \!\!-\!i,\pm\!n} &=& 
 \frac{\omega_{\scriptscriptstyle \!\pm\! i}}{\pm n/R + m_i} \;\;\;\; ,
\eeqa
where we have used the convenient notations of the original states labeled
by the index-sequence $\{+1,+2,-1,-2,+n,-n\}$ (the first four states are not to
be confused with the KK states with $n=\pm1\;\mbox{or}\; 2$)  and
$\Omega = \left(\begin{array}{cc} \omega_{\scriptscriptstyle \!+\!1} & 
\omega_{\scriptscriptstyle \!+\!2}   \\ \omega_{\scriptscriptstyle \!-\!1} & 
\omega_{\scriptscriptstyle \!-\!2} \end{array} \right)$. Perturbations to masses of the 
four light states are given by
\beqa
\Delta m_{\pm i}  &=& \sum_n  
\frac{- \omega_{\scriptscriptstyle \!+\!i}^2}{n/R \mp m_i}
+ \sum_n \frac{- \omega_{\scriptscriptstyle \!-\!i}^2}{-n/R \mp m_i} \; 
\nonumber \\
&\simeq& \mp m_i R^2 
\left( \omega_{\scriptscriptstyle \!-\!i}^2 + \omega_{\scriptscriptstyle \!+\!i}^2 \right)
\sum_n  \frac{1}{n^2}  \; .
\eeqa

\section{Generalization to Higher dimensions}
The case with one ``large" extra dimension is known to be unrealistic\cite{ADD,ADD2}.
For $\Psi$ living in $4+\delta$ dimensions with $\delta =2 $ or $3$, the gamma
matrices are $8\times 8$. There is a Weyl 4-spinor for $\delta =2 $, while
the smaller spinor representation, for $\delta =3$ or $4$, is an 8-spinor. 
Nevertheless, complications of the sort do not change the general pattern. The
higher dimensional spinor $\Psi$ can always be decomposed into KK towers of 4-D 
2-spinors with Dirac mass couplings as in the 5-D case above. Assuming the size
of the extra dimensions are all the same, the major difference is that the KK towers
now form a $\delta$-dimensional integral lattice, and we get four 2-spinors for
$\delta =3 $ or $4$ and eight 2-spinors for $\delta =5 $ or $6$.
Hence for $\delta \geq 3$,
one $\Psi$ singlet can already give masses and mixings to all three SM neutrinos.
For the purpose of the analysis in this letter, we will stick to the two 2-spinor
formulation discussed above. Generalization to the other cases is straight forward
though a bit tedious.

With $\delta$ extra dimensions, $n$ in our compressed notation should be generalized to 
mean the magnitude of a vector within the $\delta$-dimensional integral lattice,
\[
n=\sqrt{n_{\scriptscriptstyle 1}^2 + ......\;\; + 
n_{\scriptscriptstyle \delta}^2} \; ;
\] 
there would be a KK mode, with the corresponding column and row entries to
$\cal{M}$, for each of such vectors, and $\sum_n$ would be a summation over all of them.
A $\sum_n \frac{1}{n^2}$ is then always divergent. However, if the KK tower is truncated
at an ultraviolet cutoff $\sim M_*$, a sum of the form $\sum_n \frac{m^2R^2}{n^2}$,
when approximated by an integral over the $n$-lattice, goes like 
\[
\frac{S_{\delta}}{\delta -2}\; \frac{\lambda_\nu^2 v^2}{2M_*^2} =
\frac{2\pi^{\delta/2}}{\Gamma(\delta/2)}\frac{1}{(\delta -2)}\; 
\frac{\lambda_\nu^2 v^2}{2M_*^2} \; ,
\]
where $S_{\delta}$ is the result of a $\delta$-dimensional angular integration
and $v=246\;\mbox{GeV}$. The result is always $\ll 1$. Recall that
\beq \label{mv}
m = \frac{\lambda_\nu v}{\sqrt{2}} \frac{M_*}{M_{Pl}} \; ,
\eeq
and 
\beq \label{R}
R^\delta \simeq M_{Pl}^2 / M_*^{\delta +2} \; .
\eeq
Hence, adopting the truncation procedure, the validity of the perturbative results 
above is well justified.

We want to have a word of caution here about the KK tower truncation.
The general theory of ``large" extra dimensions presumes a  fundamental scale
$M_*$ and hence an ultraviolet cut-off, which has been generally taken by previous 
authors as implying a truncation of the KK tower of states. However,
in the situation 
discussed here, where the higher dimensional $\Psi$ has mass mixing with the 4-D
fields (neutrinos), it is not {\it a priori} clear whether the truncation should
rather not be applied to the physical states. The truncation of the KK tower in $\Psi$, 
and hence the mass matrix $\cal{M}$ before diagonalization, is however crucial for 
the above perturbation
results to make sense. Without the truncation, the mass eigenvalue perturbations
have $\sum_n 1/n^2$ divergences. In fact, the infinite number of $n$ states would
impose an infinite normalization factor on the matrix elements 
$U_{\scriptscriptstyle \!\!\pm\!i,\pm\!n}$ 
which is neglected in Eq.(\ref{U}). The normalizations then make each non-zero
elements in $U$ infinitesimally smaller. The physical consequence of the latter scenario
is stunning.  Any SM field localized in 4-D having mass mixing with higher 
dimensional fields will be a linear combination of an infinite number of mass
eigenstates most of them lying beyond the cut-off scale and truncated from the 
physical spectrum; hence, any SM coupling to the field has effectively vanishing
couplings to full physical spectrum of states involving the field. In particular,
applying to the neutrino mixing scenario, leptonic $W$-boson decays would be
vanishing as effective total amplitude into a physical lepton together with any physical
``neutrino" within the threshold is infinitesimally small. 

\section{Constraint from violation of charged current universality}
With the word of caution above, we go on in this section to discuss 
constraint from the violation of charged current universality as a result of the
neutrino masses and mixings, taking the usual assumption of KK tower 
truncation. We will concentrate on charged pion decay through 
$\Gamma(\pi \to e \nu)/\Gamma(\pi \to \mu \nu)$. 
The analysis involved in the case is relatively straight forward while the process
is likely to provide the most stringent constraint due to the good experimental
results and that the small $\pi$ mass makes the quantity extremely sensitive to heavy
neutrino states involved. Ref.\cite{FP} has presented such a discussion, which
we, however, do not consider to be quite complete as the effect of the
heavier,
KK tower states were not incorporated.  Though we stick here to the more
general formulation
with both types of terms in Eq.(\ref{Sint}) admitted, our result holds
even when only $\psi_{\scriptscriptstyle\! R}$ coupling is allowed, which
is the case discussed in Ref.\cite{FP} following Ref.\cite{ADDM}. 
For more background information and related
analysis of charged current universality, readers are referred to Ref.\cite{ccu}.

As discussed above, each SM neutrino $\nu_{\scriptscriptstyle\! L}$ is now given by
a linear combination of various mass eigenstates; it is predominately an equal 
mixture of a Dirac pair of mass about $\pm m_i$, with small admixture from the 
heavier physical states of masses about $\pm n/R$. We have, for charged lepton
$\ell_i$,
\small 
\beqa \label{first}
\Gamma(\pi \to \ell_i \nu) &\propto & \frac{1}{2} \Big[
\Big(1-\sum_n  |U_{\scriptscriptstyle \!\!+\!i,\!+\!n}|^2 
- \sum_n  |U_{\scriptscriptstyle \!\!+\!i,\!-\!n}|^2\Big) P^{\pi \ell_i}_{0} +
\sum_n  \Big( |U_{\scriptscriptstyle \!\!+\!i,\!+\!n}|^2
+ \sum_n  |U_{\scriptscriptstyle \!\!-\!i,\!-\!n}|^2 \Big)
 P^{\pi \ell_i}_{n} \nonumber \\ 
&&+  \Big(1-\sum_n  |U_{\scriptscriptstyle \!\!-\!i,\!+\!n}|^2 
- \sum_n  |U_{\scriptscriptstyle \!\!-\!i,\!-\!n}|^2 \Big)P^{\pi \ell_i}_{0} 
+ \sum_n \Big( |U_{\scriptscriptstyle \!\!-\!i,\!+\!n}|^2
+ \sum_n  |U_{\scriptscriptstyle \!\!-\!i,\!-\!n}|^2 \Big)
 P^{\pi \ell_i}_{n}  \Big] \; ,
\eeqa 
\normalsize
where 
\beq
P^{\pi \ell_i}_{n} = 
\theta(m_\pi-m_{\ell_i}-m_{\nu_{\scriptscriptstyle n}})\,
\left[\frac{m_{\ell_i}^2}{m_\pi^2} + \frac{m_{\nu_{\scriptscriptstyle n}}^2}{m_\pi^2} -
\left(\frac{m_{\ell_i}^2}{m_\pi^2} 
- \frac{m_{\nu_{\scriptscriptstyle n}}^2}{m_\pi^2}\right)^2 \right] \;\;
\lambda^{1/2}\!\left(1,\frac{m_{\ell_i}^2}{m_\pi^2},
\frac{m_{\nu_{\scriptscriptstyle n}}^2}{m_\pi^2}\right) \; ,
\eeq
is the matrix element-phase space function, with $\lambda(a,b,c) = (a-b-c)^2 - 4bc$,
and $m_{\nu_{n}}^2 \simeq n^2/R^2$ being the square of the physical neutrino masses;
$P^{\pi \ell_i}_{0}$ is a bit of abuse of notation as the physical mass $\pm m_i$
is safely approximated as zero. We have shown in the above formula only the dependence on
the partial width of the masses and mixings. Should all the physical neutrino masses be 
negligible, there would, of course, be no change in the decay width. The important point to
note here is that the KK tower contributes to arbitrarily heavy
neutrino states up to the
ultraviolet cutoff beyond the $\pi$-decay threshold. We rewrite the expression
in the form
\beq 
\Gamma(\pi \to \ell_i \nu) \propto  P^{\pi \ell_i}_{0}
\left(1-  \sum_n^{\scriptscriptstyle (\!\pi\! -\!\ell_{\!i}\! )\!'} 
\frac{(\omega_{\scriptscriptstyle +\!i}^2 
+ \omega_{\scriptscriptstyle -\!i}^2 )R^2}{n^2}
+  \sum_n^{\scriptscriptstyle \!\pi\! -\!\ell_{\!i}} 
\frac{(\omega_{\scriptscriptstyle +\!i}^2 
+ \omega_{\scriptscriptstyle -\!i}^2 )R^2}{n^2}
{\cal P}^{\pi \ell_i}_{\!n} \right) \; ,
\eeq
where we have substituted Eq.(\ref{U}) with $m_i$ neglected, and introduced
\beq
{\cal P}^{\pi \ell_i}_{\!n} = \frac {P^{\pi \ell_i}_{n}-P^{\pi \ell_i}_{0}}
{P^{\pi \ell_i}_{0}} \; ;
\eeq
whereas the summation $\sum_n^{\scriptscriptstyle \!\pi\! -\!\ell_{\!i}} $ 
indicates a sum up to the $\pi$-decay threshold,
{\it i.e.} $|n| \leq R(m_\pi - m_{\ell_i})$ and the summation 
$\sum_n^{\scriptscriptstyle (\!\pi\! -\!\ell_{\!i}\! )\!'}$ goes 
from there to the ultraviolet cutoff $|n| \leq RM_*$. We further simplify the 
expression to
\beq \label{pi-l}
\Gamma(\pi \to \ell_i \nu) \propto  P^{\pi \ell_i}_{0}
\left[ 1 + (\lambda_{\scriptscriptstyle +\!i}^2 
+ \lambda_{\scriptscriptstyle -\!i}^2 )
\frac{R^2 v^2 M_*^2}{2 M_{Pl}^2}
\left(\sum_n^{\scriptscriptstyle \!\pi\! -\!\ell_{\!i}}
 \frac{1}{n^2} {\cal P}^{\pi \ell_i}_{\!n} -
\sum_n^{\scriptscriptstyle (\!\pi\! -\!\ell_{\!i}\! )\!'} 
\frac{1}{n^2} \right) \right] \; ,
\eeq
where we introduced, in accordance with Eq.(\ref{mv}),
\beq
\omega_{\scriptscriptstyle \!\pm\!i} = \frac{\lambda_{\scriptscriptstyle \pm\!i}v}{\sqrt{2}} 
\frac{M_*}{M_{Pl}} \; .
\eeq
Here, $\lambda_{\scriptscriptstyle \pm\!i}$'s can be considered as Yukawa couplings
in some chosen basis.

In the above expressions, we have implicitly assumed the charged leptons $\ell_i$'s are
physical states. Actually, there should be a leptonic-CKM matrix
involved as well.
 However, to the extent that we have no information about the original
Yukawa couplings ($\lambda$'s) as given in Eqs.(\ref{Sint}) and
(\ref{mv}), we can do no better than stating that Eq.(\ref{pi-l}) applies to $e$ and
$\mu$ each with  
$(\lambda_{\scriptscriptstyle +\!i}^2 + \lambda_{\scriptscriptstyle -\!i}^2 )$
replaced by a corresponding quantity dependent on the original Yukawa couplings.
Hence, we write
\beq \label{final}
\frac{\Gamma(\pi \to e \nu)}{\Gamma(\pi \to \mu \nu)} =
R^{\pi e}_{\pi \mu}
\frac{\left[ 1 + Y_e \frac{R^2 v^2 M_*^2}{2 M_{Pl}^2} \left(
\sum_n^{\scriptscriptstyle \!\pi\! -\!e}
 \frac{1}{n^2} {\cal P}^{\pi e}_{\!n} -
\sum_n^{\scriptscriptstyle (\!\pi\! -\!e)\!'} 
\frac{1}{n^2} \right) \right]}
{\left[ 1 + Y_\mu \frac{R^2 v^2 M_*^2}{2 M_{Pl}^2}  \left(
\sum_n^{\scriptscriptstyle \!\pi\! -\!\mu}
 \frac{1}{n^2} {\cal P}^{\pi \mu}_{\!n} -
\sum_n^{\scriptscriptstyle (\!\pi\! -\!\mu)\!'} 
\frac{1}{n^2} \right) \right]} \; ,
\eeq
where $Y_e$ and $Y_\mu$, of the same order as the $\lambda^2$'s,
parametrize our ignorance about the Yukawa couplings as pointed
out above, and
\beq
R^{\pi e}_{\pi \mu}   = 1.233 \times 10^{-4}
\eeq
is the SM result. The experimental number is $(1.230\pm0.004)\times 10^{-4}$
at one $\sigma$. With only $\psi_{\scriptscriptstyle \! R}$ allowed to
coupled, an extra $\psi_{\scriptscriptstyle \! R}$ has to be used to
replace $\psi_{\scriptscriptstyle \! L}^c$, but the formula here for
$\frac{\Gamma(\pi \to e \nu)}{\Gamma(\pi \to \mu \nu)}$ applies unchanged.

To check the explicit modification of the $\pi$-decay ratio, and hence the corresponding
constraint on the ``large" extra dimension scenario, we evaluate the relevant
terms in the above equation, approximating the summation by an integration. Some numerical
results are listed in Table 1. Reading from the table, we can draw lower bounds on
$M_*$ using the experimental result, for various number of extra dimensions. The kind of 
bounds obtained gives only an order of magnitude. The first reason for this is
our ignorance in $Y_e$ and $Y_\mu$, which potentially can be improved by other
complimentary analysis such as fitting neutrino oscillation data. A deeper problem comes
from our ignorance of the ultraviolet regime. We have used a simple cutoff at $M_*$
to truncate the KK tower. In the numerical calculation, we have implemented 
Eq.(\ref{R}) with an equality, which in term can be considered as our definition
of $M_*$ itself. However, the numerical result is not too sensitive to the explicit 
cutoff value used; the $\sum_n^{\scriptscriptstyle \!\pi\! -\!\ell}$ terms are
not dependent on it while the  $\sum_n^{\scriptscriptstyle (\!\pi\! -\!\ell)\!'}$ terms 
are dominated by small $n$. For example, we have checked explicitly that changing the cutoff by an 
order does not change our numbers in any substantial way.

As shown in Table 1, the  $\sum_n^{\scriptscriptstyle \!\pi\! -\!\mu}
 \frac{1}{n^2} {\cal P}^{\pi \mu}_{\!n}$ term is the smallest one in Eq.(\ref{final}),
never playing a significant role. This is mainly a result of the very small threshold, and
the generic small value of ${\cal P}^{\pi \mu}_{\!n}$. On the contrary,
${\cal P}^{\pi e}_{\!n}$ can reach very large value for $n/R$ (neutrino mass) in the
MeV region. There is a strong suppression from the 
$ |U_{\scriptscriptstyle \!\!\pm\!i,\!\pm\!n}|^2 $ prefactors [{\it cf.} Eq(\ref{first})],
but also a huge number of relevant states contributing. For $\delta=2$, the 
$\sum_n^{\scriptscriptstyle \!\pi\! -\!e} \frac{1}{n^2} {\cal P}^{\pi e}_{\!n}$ 
term dominates and gives large modifications to ${\Gamma(\pi \to e \nu)}$. In that case,
the numerical value of the term scales as $M_*^{\mbox{-}\delta}$, to a 
very good approximation. From the table, we can see that this implies 
a lower bound of roughly $M_*<1600\mbox{TeV}$, assuming $Y_e$ to be order one.
For $\delta>2$, however, $1/R$ increases pretty fast which results in a strong
suppression in the term. The two $\sum_n^{\scriptscriptstyle (\!\pi\! -\!\ell)\!'}$
terms go as $M_*^{\mbox{-}2}$ for all $\delta>2$ and become dominating. 
The two terms tend to cancel one another, with final result 
dependent on $Y_e -Y_\mu$. Taking the latter
to be order one, with $Y_\mu>Y_e$, gives roughly $M_*<20\mbox{TeV}$ with a very weak
dependence on $\delta$. In fact, 
\beq
\frac{R^2 v^2 M_*^2}{2 M_{Pl}^2}
\sum_n^{\scriptscriptstyle (\!\pi\! -\!\ell)\!'}\frac{1}{n^2}
\simeq \frac{\pi^{\delta/2}}{\Gamma(\delta/2)}\frac{1}{(\delta -2)}\; 
\frac{v^2}{M_*^2} \; ,
\eeq
when the lower limits of summation are neglected, which serves as a good 
approximation for all the $\delta\leq 3$ cases as suggested by the apparent
identical results in the $\sum_n^{\scriptscriptstyle (\!\pi\! -\!e)\!'}$ and
$\sum_n^{\scriptscriptstyle (\!\pi\! -\!\mu)\!'}$ ({\it cf.} third and last columns
of Table I). Hence we refrain from presenting explicit number for $\delta>4$.

A word of caution is in order for the interpretation of the results. 
While taking the $\lambda$-type couplings
to be not substantially smaller than unity is a common
strategy among most previous authors on the subject, a suppression in
their values would weaken the constraints obtained very effectively.
As stated above, the constraints go as the squares of the $\lambda$'s.
Hence, suppressed $\lambda$ values would easily render the constraints 
obtained much less interesting. From the theoretical point of view,
assuming order one couplings avoids the need of an extra mechanism to produce 
the otherwise suppressed values. Hence, the strategy is commonly adopted.  Phenomenologically speaking, however, one 
would have to check if there are numbers that can fit in with all the 
experimental data, through a comprehensive analysis of all aspects of neutrino
physics from a specific model. That is a task beyond the present work. 

\section{Concluding remarks}
Neutrino physics is rather subtle and unconventional  in
a theory with extra dimensions not seen by the other SM fields. While 
singlet fermion(s) living in at least some of the extra dimension(s) and
coupling to the SM neutrinos seems to provide a mechanism to naturally
explain the small neutrino masses, a comprehensive analysis of all aspects
of neutrino phenomenology still needs to be performed. We have
discussed  here only some 
of the issues involved and presented numerical results from the pion decay
constraints. Assuming order one Yukawa couplings, our results give a
$1600\;\mbox{TeV}$ bound on $M_*$
for two extra dimensions, substantially stronger than any other known 
constraints. The high $M_*$ value also implies larger neutrino mass for the lightest
states, threatening fittings with neutrino oscillations\cite{osc}. For more than
two extra dimensions, the bound stays around $20\;\mbox{TeV}$.
It should be emphasized again that the above strong numerical bounds
are based on order one Yukawa couplings, which is pushing the
limit of validity of the perturbational analysis used. More
realistic couplings, say 0.3, actually give a bound no stronger 
than the existing bounds which can be found in Ref.\cite{ascos}.

The bounds obtained here do not necessarily rule out smaller $M_*$,
however, if  an alternate neutrino mass generation mechanism is assumed.
Results would also be modified if the singlet fermion(s) sees some, instead of
all, of the extra dimensions that gravity sees. In any case, we expect various
charged current universality constraints to be very important, compared with
other more direct collider bounds. Neutrino physics may provide the 
best probe into the extra dimensions, if they exist.

We would like to thank G. Shiu and G.-H. Wu for discussions.
This work was supported in part by the U.S. Department of Energy,
under grant DE-FG02-91ER40685


\clearpage


{{\bf Table Caption}}:-

\vskip 2.0cm
\noindent
{\bf Table I}:
Numerical Results on terms contributing to deviation of
$\frac{\Gamma(\pi \to e \nu)}{\Gamma(\pi \to \mu \nu)}$ from the Standard Model
value, as given by Eq.(\ref{final}).

\bigskip
\bigskip

\vspace*{1in}

\hrule

\vspace*{1in}
\noindent
{\bf Table I :-}\\

\noindent
\begin{tabular}{||cc||c|c|c|c||}\hline \hline
$\delta$ & $M_*\;(\mbox{TeV})$ & 
$\frac{R^2 v^2 M_*^2}{2 M_{Pl}^2}  \sum_n^{\scriptscriptstyle \!\pi\! -\!e}
 \frac{1}{n^2} {\cal P}^{\pi e}_{\!n}$ & 
$\frac{R^2 v^2 M_*^2}{2 M_{Pl}^2}  \sum_n^{\scriptscriptstyle (\!\pi\! -\!e)\!'} 
\frac{1}{n^2}$ &
$\frac{R^2 v^2 M_*^2}{2 M_{Pl}^2}  \sum_n^{\scriptscriptstyle \!\pi\! -\!\mu}
 \frac{1}{n^2} {\cal P}^{\pi \mu}_{\!n}$ & 
$\frac{R^2 v^2 M_*^2}{2 M_{Pl}^2}  \sum_n^{\scriptscriptstyle (\!\pi\! -\!\mu)\!'} 
\frac{1}{n^2}$ \\ \hline
$2$ & $1$       & $7.52\pi\times 10^{2}$        & $5.37\pi\times 10^{-1}$
                & $-6.74\pi\times 10^{-3}$      & $6.23\pi\times 10^{-1}$
\\
        & $100$ & $7.52\pi\times 10^{-2}$       & $8.16\pi\times 10^{-5}$
                & $-8.47\pi\times 10^{-7}$      & $9.02\pi\times 10^{-5}$
\\
        & $1600$ & $2.94\pi\times 10^{-4}$      & $3.84\pi\times 10^{-7}$
                & $-3.31\pi\times 10^{-9}$      & $4.18\pi\times 10^{-7}$
\\ \hline
$3$ & $1$       & $9.60\pi\times 10^{-2}$       & $1.21\pi\times 10^{-1}$
                & $-4.79\pi\times 10^{-7}$      & $1.21\pi\times 10^{-1}$
\\
        & $10$  & $9.60\pi\times 10^{-5}$       & $1.21\pi\times 10^{-3}$
                & $-4.79\pi\times 10^{-10}$     & $1.21\pi\times 10^{-3}$
\\
        & $20$  & $1.20\pi\times 10^{-5}$       & $3.03\pi\times 10^{-4}$
                & $-5.99\pi\times 10^{-11}$     & $3.03\pi\times 10^{-4}$
\\ \hline
$4$ & $1$       & $3.66\pi^2\times 10^{-6}$     & $3.03\pi^2\times 10^{-2}$
                & $-6.99\pi^2\times 10^{-12}$   & $3.03\pi^2\times 10^{-2}$
\\
        & $10$  & $3.66\pi^2\times 10^{-10}$    & $3.03\pi^2\times 10^{-4}$
                & $-6.99\pi^2\times 10^{-16}$   & $3.03\pi^2\times 10^{-4}$
\\
        & $20$  & $2.29\pi^2\times 10^{-11}$    & $7.56\pi^2\times 10^{-5}$
                & $-4.37\pi^2\times 10^{-17}$   & $7.56\pi^2\times 10^{-5}$
\\ \hline
\hline
\end{tabular}

\end{document}